\documentclass[aps,prb,twocolumn,superscriptaddress,groupedaddress,showpacs,tightenlines,balancelastpage,raggedbottom]{revtex4-1}  
\usepackage{graphicx}
\usepackage{dcolumn}
\usepackage{sidecap}
\usepackage{float}
\usepackage{bm}
\usepackage{amssymb}
\usepackage{amsmath}
\usepackage{subfigure}
\usepackage{color}
\usepackage[a4paper]{geometry}
\newgeometry{top=2cm,right=1.25cm,left=1.25cm,bottom=2cm}

\begin{document}

\widetext
\title{Electronic and Optical Properties of [110]-Tilted InAs/GaAs Quantum Dot Stacks}


\author{Muhammad Usman}
\email{usman@alumni.purdue.edu}
\affiliation{Tyndall National Institute, Lee Maltings, Dyke Parade, Cork, Ireland}

\vskip 0.25cm



\begin{abstract}

Multi-million atom simulations are performed to study stacking-angle ($\theta$) dependent strain profiles, electronic structure, and polarization-resolved optical modes from [110]-tilted quantum dot stacks (QDSs). Our calculations reveal highly asymmetrical biaxial strain distributions for the tilted QDSs that strongly influence the confinements of hole wave functions and thereby control the polarization response. The calculated values of degree of polarizations, in good agreement with the available experimental data, predict a unique property of the tilted QDSs that the isotropic polarization response can be realized from both [110] and [-110] cleaved-edges $-$ a feature inaccessible from the conventional [001]-QDSs. Detailed investigations of polar plots further establish that tilting the QDSs provides an additional knob to fine tune their polarization properties.

\end{abstract}

\pacs{81.07.Ta, 81.05.Ea, 78.55.Kz, 81.15.Hi}

\maketitle

Since the experiment by Xie \textit{et al}.\cite{Xie_1} demonstrating strain-driven vertical self-organization of InAs/GaAs quantum dots in which the vertically correlated quantum dot layers (QDLs) aligned along the growth axis to form quantum dot stacks (QDSs), such nano-structures have been a topic of extensive research due to their promising electronic and optical properties for implementing several optoelectronic devices such as semiconductor optical amplifiers (SOAs)\cite{Takahashi_1, Alonso_1, Inoue_1, Usman_1, Usman_7, Ikeuchi_1}, lasers diodes\cite{Bimberg_2, QDM_1}, infrared photo-detector\cite{Kim_1}, high efficiency intermediate-band solar cells (IBSC)\cite{Oshima_1, Luque_1}, etc. Conventionally the strongly coupled QDSs grow in the growth direction with the constituent QDLs well aligned along the [001]-axis, and their theoretical understanding is well established in the literature\cite{Usman_1, Saito_1, Andrzejewski_1, QDM_1, Ridha_1}. However it is only very recently that the growth of tilted QDSs is reported~\cite{Bessho_1}, where the stacking direction is controllable by varying the direction of In-flux during the self-assembly growth process. This in turn led to the formation of QD stacks tilted towards the [110]-direction, with the angle of tilt with respect to the [001]-direction closely related to the incident angle of the In-flux.    

The polarized PL measurements on the tilted QDSs, without any theoretical guidance, displayed polarization properties drastically different from the conventional [001]-aligned QDSs. For example, TE$_{[110]} <$ TM$_{[001]}$ for the [001]-aligned stack~\cite{Inoue_1, Ikeuchi_1} changed to TE$_{[110]} >$ TM$_{[001]}$ for the tilted stack\citep{Bessho_1} for the same number of QDLs in both stacks. Furthermore, a large rotation of PL polar plot peak was measured from the [-110] cleaved-edge surface leading to a discrepancy with the angle of tilt of the QDS. This appealed for a detailed understanding of the stacking-angle dependent electronic and optical properties to appropriately explain the experimental measurements.         

\begin{figure}
\includegraphics[scale=0.35]{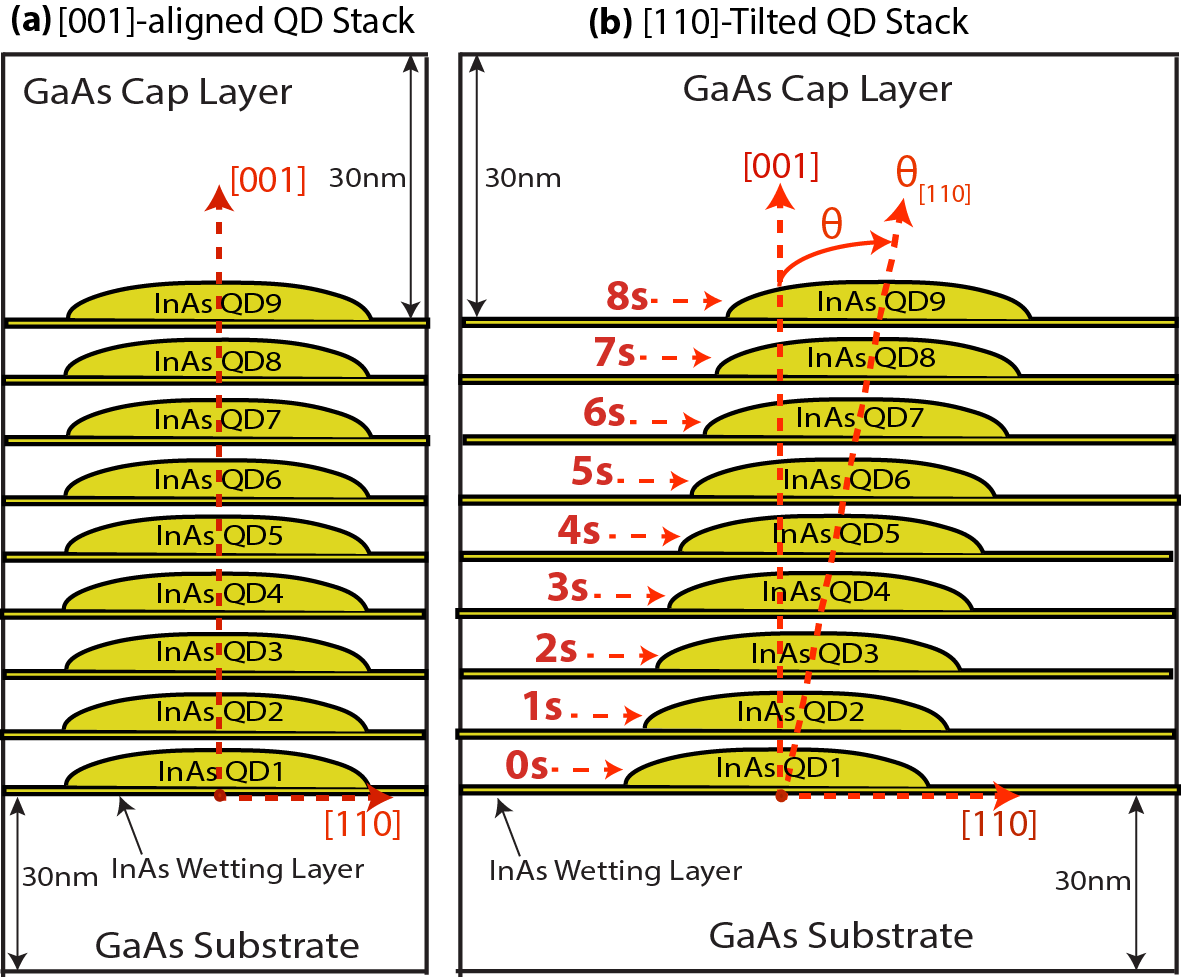}
\caption{Schematics of the two stacks consisting of nine QD layers are shown where: (a) the QD layers are aligned along the [001]-direction. (b) the QD layers are tilted towards the [110]-direction by an angle $\theta$ with respect to the [001]-direction.}
\label{fig:Fig1}
\end{figure}

This work presents a first theoretical study of the tilted InAs QDSs by performing atomistic simulations of strain, electronic structure, and polarization-dependent inter-band optical transition strengths. The unique characteristics of the tilted QDSs are reported by highlighting highly asymmetrical biaxial strain distributions that regulate the electronic structure and polarization properties. It is predicted that by tilting the QDS, isotropic polarization response could be simultaneously realized from both [110] and [-110] cleaved-edge surfaces $-$ a feature not accessible from the conventional [001]-aligned stacks\cite{Ikeuchi_1, Usman_1, Usman_7}. The comparison of polar plots further revealed a high degree of control over the polarization properties attainable by tilting the QDSs, making it a useful tool for engineering the optical properties from the QD nano-structures.     

Fig.~\ref{fig:Fig1} illustrates schematic diagrams of two QDSs consisting of nine QDLs (labelled as QD1 to QD9 from the bottom to the top) with: (a) the centers of all the QDLs perfectly aligned along the [001] direction. This stack will be referred to as [001]-aligned QDS. and (b) the center of $n^{th}$ QDL shifted by $s(n-1)$ (nm) in the [110]-direction. This type of QDS will be called [110]-tilted QDS. The angle of tilt ($\theta$) of the stack with respect to the [001]-direction is directly related to the value of $s$ which has been selected as multiples of $\sqrt{2} a$. It should be noted that a shift of $\sqrt{2} a$ in a QDL center along the [110] direction implies shifting the center simultaneously in both [100] and [010] directions by $a$, where $a$ is the GaAs lattice constant. The various values of $s$ and the corresponding values of $\theta$ are provided in table~\ref{tab:Angle}. Clearly selecting $s$ = 0 leads to a [001]-aligned stack with $\theta$ = 0. Finally, we have selected the geometry parameters of the QDLs (Shape = Dome, Base diameter = 20 nm, Height = 4 nm, wetting-layer thickness = $a$ nm, and spacer thickness = 9$a$ nm) as reported by the experiments~\cite{Inoue_1, Bessho_1} so that to enable a direct comparison of our calculated values with the measured values. 

\begin{table}[tb]
\caption{\label{tab:Angle} The values of the tilt angle ($\theta$) and the shift in the centers of the QDLs along the [110]-direction ($s$) are provided; $a$ is the unstrained lattice constant of GaAs.}
\begin{ruledtabular}
\begin{tabular}{ccccccccc}
    $s$ (nm) & 0 & $\sqrt{2} a$ & 2$\sqrt{2} a$ & 3$\sqrt{2} a$ & 4$\sqrt{2} a$ & 5$\sqrt{2} a$  \\
    \hline
    $\theta$ (degrees) & 0 & 8$^{o}$ & 15.8$^{o}$ & 23$^{o}$ & 29.5$^{o}$ & 35.3$^{o}$  \\
\end{tabular}
\end{ruledtabular}
\end{table}

The simulations are performed using atomistic tool NEMO 3-D~\cite{Klimeck_1, Klimeck_2}, where the strain is computed from the valence force field (VFF) relaxation of atoms~\cite{Keating_1, Olga_1} and the electron and hole energies and states are obtained by solving the twenty-band sp$^3$d$^5$s* tight-binding Hamiltonian~\cite{Boykin_1}. The polarization-resolved inter-band optical transition strengths are calculated from Fermi's Golden Rule and summing the absolute values of the momentum matrix elements over the spin degenerate states~\cite{Usman_1, Graf_1, Boykin_2001, Boykin_Vogl}. A very large simulation domain consisting of up to 43 million atoms is used to properly accommodate the effects of long range strain fields. The GaAs box surrounding the QDSs have realistic boundary conditions as reported in Ref.~\onlinecite{Usman_5} and the surface atoms are passivated according to our published recipe~\cite{Lee_1}.           
   
From our calculations, the hydrostatic strain ($\epsilon_{H} = \epsilon_{xx} + \epsilon_{yy} + \epsilon_{zz}$) is only negligibly changed when the stack is tilted. Since the electron energies are only affected by the hydrostatic strain component~\cite{Usman_3}, so we calculate a very small increase ($\leq$ 5 meV) in the lowest two electron energies (E1 and E2) when $\theta$ in increased from 0 to 35.3$^o$. At $\theta$ = 0, the lowest two electron wave functions are strongly coupled molecular wave functions with E1 being the bonding state and E2 being the anti-bonding state. However, when the QDS is tilted towards the [110]-direction, the strong coupling between the QDLs weakens, and the two electron wave functions shift in the opposite directions inside the stack: E1 becomes more localized towards the bottom of the stack and E2 tends to be more localized in the upper QDLs of the stack.      

\begin{figure}
\includegraphics[scale=0.3]{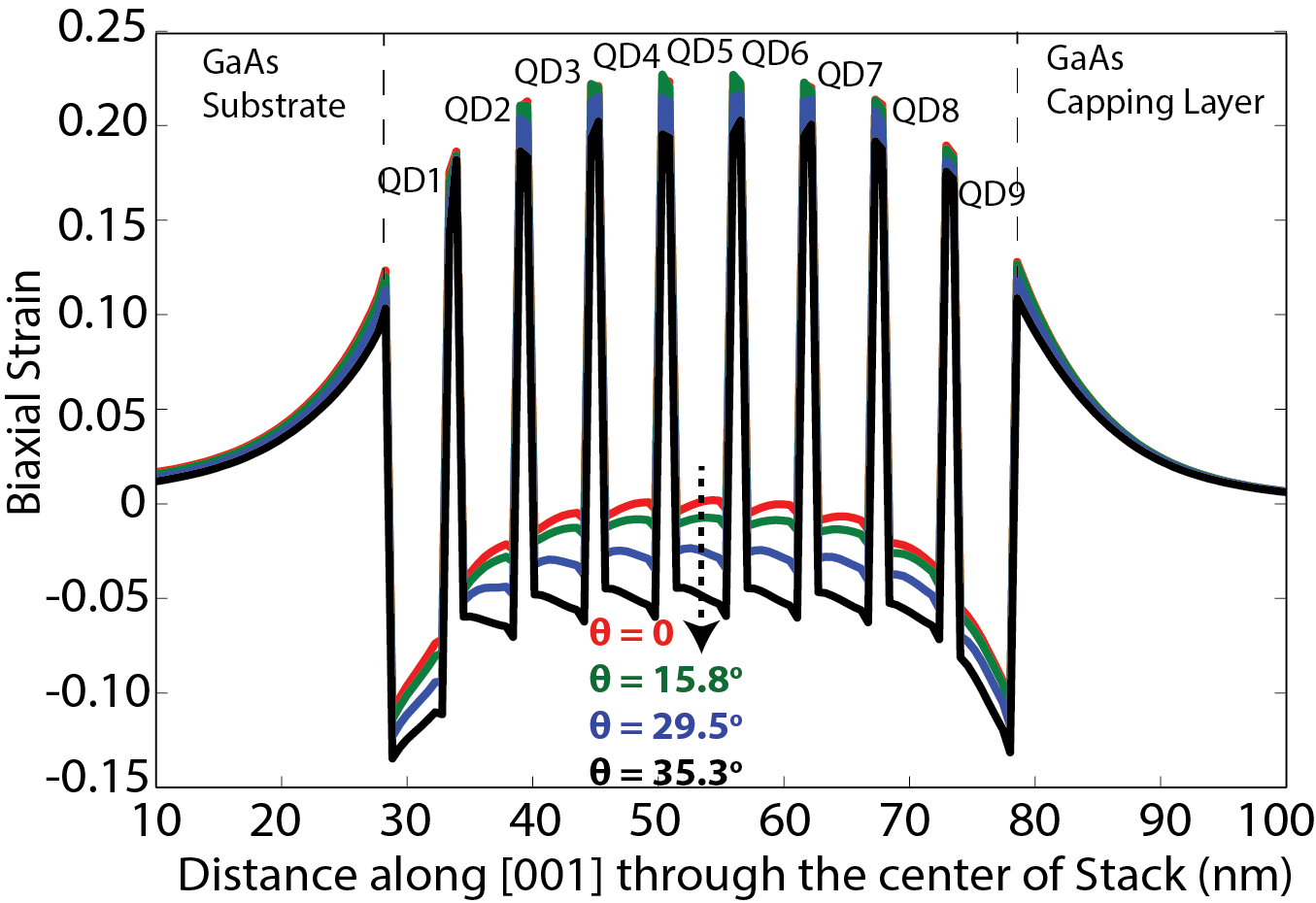}
\caption{Biaxial strain component through the centers of the stacks along the $\theta$-direction.}
\label{fig:Fig2}
\end{figure}

Contrarily, the biaxial strain ($\epsilon_{B} = \epsilon_{xx} + \epsilon_{yy} - 2\epsilon_{zz}$) is substantially affected due to the tilting of the QDS. This is evident from the 1-D plots in Fig.~\ref{fig:Fig2} where the biaxial strain is plotted through the centers of the QDSs along the $\theta$-direction. For the [001]-aligned QDS ($\theta$ = 0), the biaxial strain is significantly relaxed and becomes nearly zero at the center of the stack (see the red curve), as expected for a strongly coupled QDS~\cite{Usman_1, Saito_1}. As the QDS is tilted, the relaxation of the biaxial strain systematically reduces ascertaining that the strong coupling between the QDLs becomes weaker as a function of $\theta$. This induces an increase in the hole energies, shifting for example the highest hole energy (H1) by $\approx$38 meV when $\theta$ is changed from 0 to 35.3$^o$. 

\begin{figure}
\includegraphics[scale=0.32]{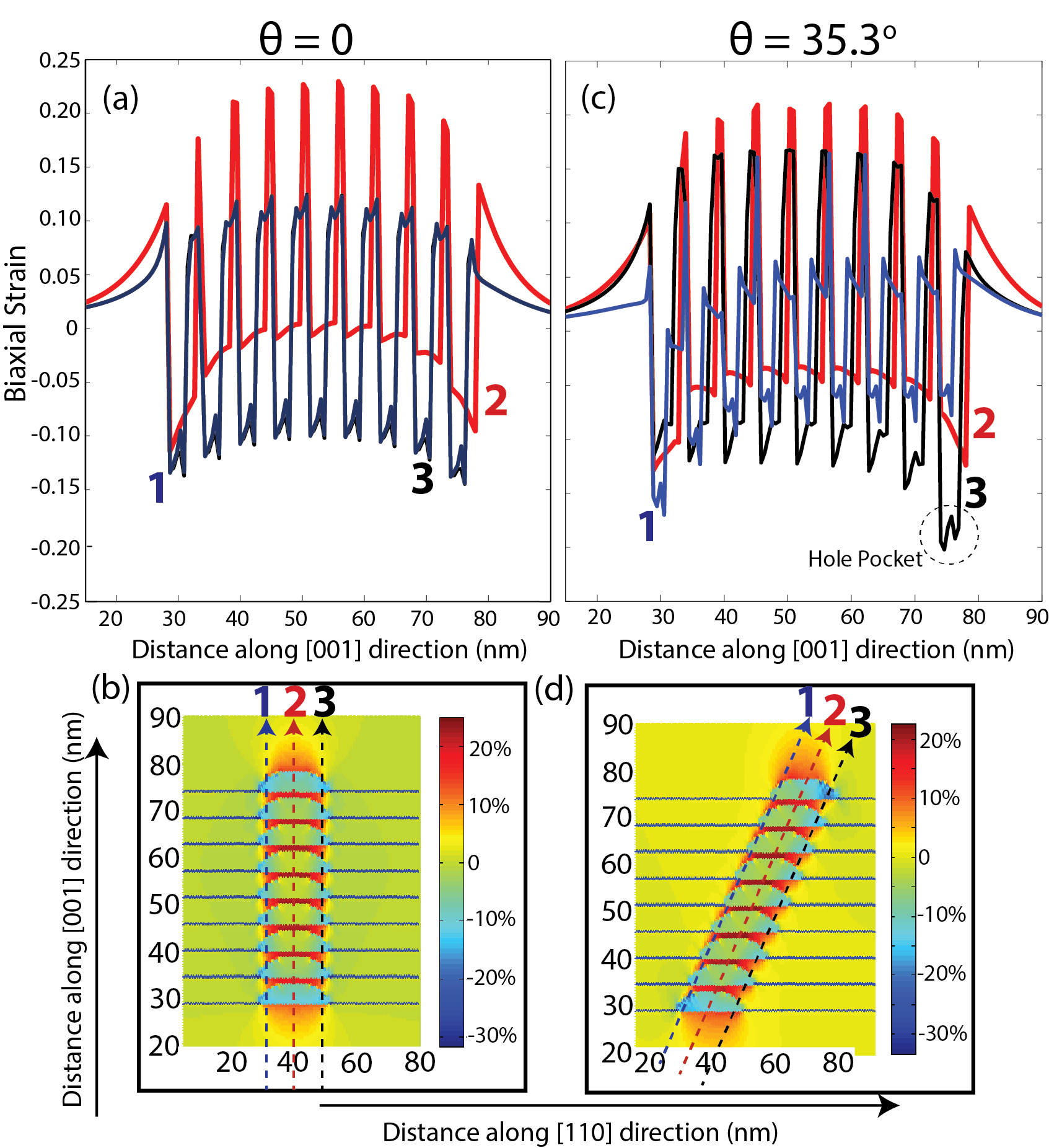}
\caption{(b, d) Biaxial strain components are plotted through the centers of the QDSs in the [110]-[001] plane for $\theta$ = 0 and 35.3$^o$. (a, c) The 1-D curves of the biaxial strain along the three lines as illustrated by (1)blue, (2)red, and (3)black arrows in (b) and (d).}
\label{fig:Fig3}
\end{figure}

A second striking effect of tilting the QDS is related to the distribution of the biaxial strain profile which significantly changes as a function of $\theta$. This effect is highlighted in Fig.~\ref{fig:Fig3} by comparing the biaxial strain distributions for the two QDSs: (a, b) [001]-aligned QDS with $\theta$ = 0 and (c, d) [110]-tilted QDS with $\theta$ = 35.3$^o$. In Figs.~\ref{fig:Fig3} (a) and (c), 1-D strain plots are shown along the three lines marked by (1) blue, (2) red, and (3) black arrows as illustrated in Figs.~\ref{fig:Fig3} (b) and (d). For the [001]-aligned QDS, we find that the biaxial strain distribution is nearly symmetric around its center (red arrow): approximately equal negative values along the two lines close to the edges (blue and black arrows) and a strong relaxation occurring along the central line (red arrow). However for the tilted QDS, the biaxial strain distribution becomes highly asymmetrical around the central red arrow, with the magnitude of the biaxial strain being considerably different along the three vertical lines. The region where the biaxial strain is relaxed has now slightly moved towards the left edges of the QDLs, with the right edges being under stronger negative biaxial strain, especially for the higher QDLs. Noticeably, the large negative biaxial strain on the right edge of the QD9 will lead to the creation of hole pocket in the QD9 layer much similar to the ones earlier observed for the single~\cite{Usman_5, Narvaez_1} and stacked~\cite{Usman_7} QDs. As will be sown next, this will lead to the confinement of hole wave functions in the QD9 layer along the [110]-direction, thereby considerably impacting the polarization properties.

\begin{figure}
\includegraphics[scale=0.16]{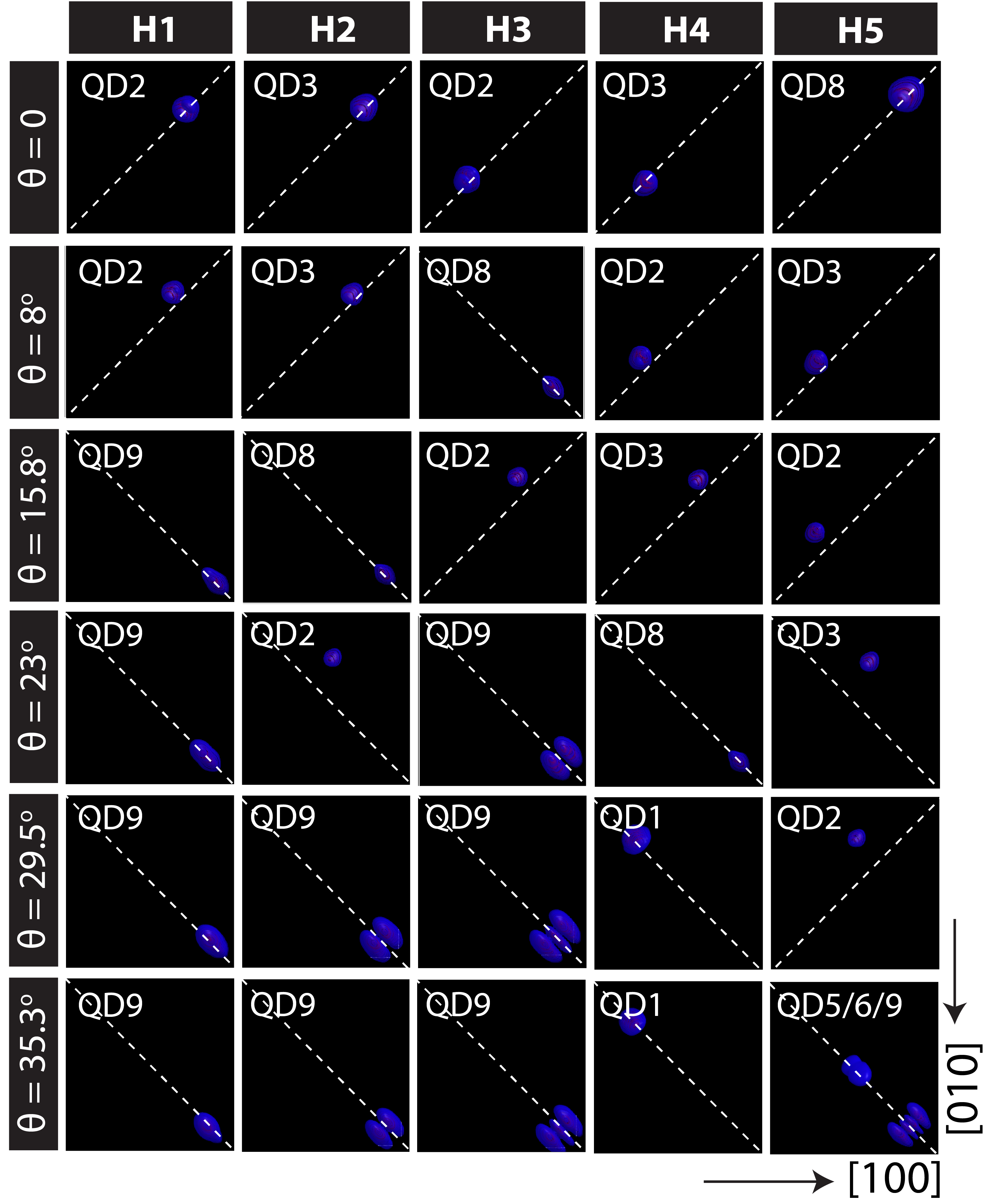}
\caption{The top views of the first five hole wave functions are shown for various values of $\theta$. Each plot indicates the orientation of a hole wave function along either [110] or [-110] axis by a dotted line and the QDL number (from QD1 to QD9) in which that particular hole wave function is spatially confined.}
\label{fig:Fig6}
\end{figure}

The hole wave functions for the highest five confined hole states (H1 $-$ H5) are plotted in Fig.~\ref{fig:Fig6} for various values of $\theta$. Each plot shows the top view of a hole wave function and also mark: (i) the orientation of that hole wave function along either [110] or [-110]-direction by a dotted line. (ii) the QDL number where that hole wave function is confined inside the stack. This is because while the electron wave functions are molecular-like states hybridized over multiple QDLs, the hole wave functions remain typically confined within the individual QDLs due to their heavier mass~\cite{Usman_1}.       

From Fig.~\ref{fig:Fig6}, all the hole wave functions are oriented along the [-110]-direction for the conventional [001]-aligned QDS as also reported in Ref.~\onlinecite{Usman_1}. When the QDS is tilted towards the [110]-direction, the wave function orientations are systematically changed from the [-110]-direction to the [110]-direction. This is mainly due to the fact that the magnitude of the negative biaxial strain increases at the [110] edges of the QDLs in the tilted stacks. For $\theta$ = 35.3$^o$, all of the highest five hole wave functions are oriented along the [110]-direction. A second noticeable change is that the hole wave functions shift in the top most QDL as $\theta$ increases. For $\theta$ = 35.3$^o$, H1, H2, H3, and H5 are confined in the top most QD9 layer due to the formation of hole pocket as discussed earlier. We also note that for $\theta$ = 35.3$^o$, the H5 state is hybridized in the QD5 and QD6 layers and this hybridization of the hole wave function resulting as a consequence of tilting the QDS would introduce a large increase in the inter-band optical modes due to an increased overlap with the hybridized electron state (E1).

\begin{figure}
\includegraphics[scale=0.35]{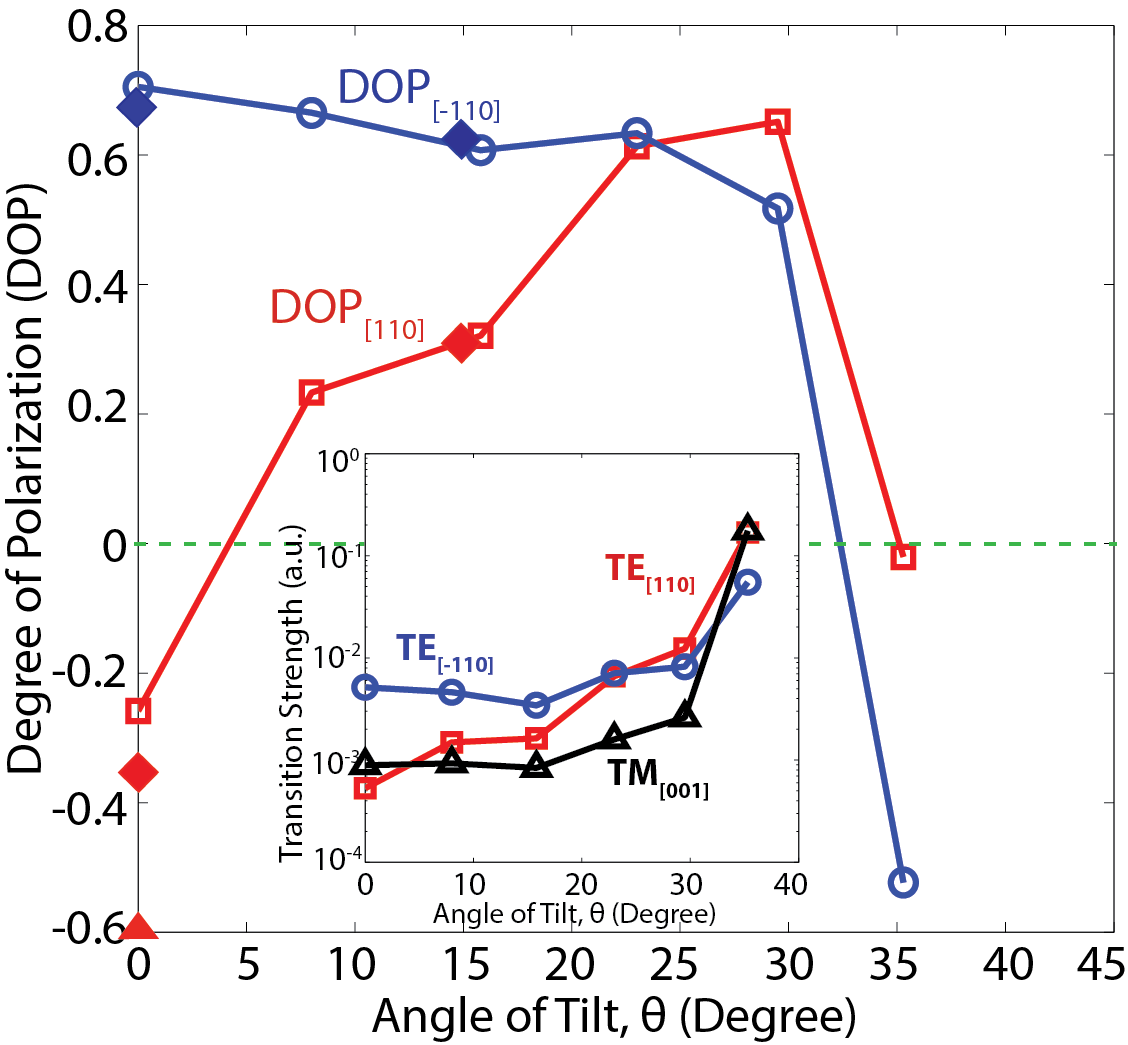}
\caption{Plots of calculated DOP$_{[110]}$ (squares) and DOP$_{[-110]}$ (circles) are shown as a function $\theta$. The available experimental values at $\theta$ = 0 (the diamond symbols from Ref.~\onlinecite{Ikeuchi_1} and a triangular symbol from Ref.~\onlinecite{Inoue_1}) and $\theta$ = 15$^o$ (the diamond symbols from Ref.~\onlinecite{Bessho_1}) are also included. Inset plots the calculated values of TE$_{[110]}$, TE$_{[-110]}$, and TM$_{[001]}$ mode transition strengths.}
\label{fig:Fig7}
\end{figure}

The polarization dependent optical properties of the QDSs are studied in Fig.~\ref{fig:Fig7}, where we plot the calculated values of DOP$_{[110]}$ = (TE$_{[110]} -$ TM$_{[001]}$)/(TE$_{[110]}$ + TM$_{[001]}$) and DOP$_{[-110]}$ = (TE$_{[-110]} -$ TM$_{[001]}$)/(TE$_{[-110]}$ + TM$_{[001]}$) as a function of $\theta$. The available experimental values~\cite{Bessho_1, Inoue_1, Ikeuchi_1} are also shown by using diamond and triangle symbols. For $\theta$ = 0, DOP$_{[110]} <$ 0 and DOP$_{[-110]} >$ 0 due to the orientations of the hole wave functions along the [-110]-direction, in agreement with the experimental measurements. When the QDS is tilted, the hole wave function orientations are changed from [-110] to [110], leading to a large increase in the TE$_{[110]}$ mode as evident in the inset of Fig.~\ref{fig:Fig7}. The changes in the TE$_{[-110]}$ and TM$_{[001]}$ modes are relatively small for 0 $< \theta <$ 35.3$^o$. As a result of these changes, the DOP$_{[110]}$ increases drastically and changes sign from negative to positive, whereas the DOP$_{[-110]}$ slightly decreases for $\theta$ between 0 and 35.3$^o$. The calculations also show a good agreement with the experimentally measured values\citep{Bessho_1} at $\theta$ = 15$^o$ confirming the accuracy of our modelling technique. At $\theta$ = 35.3$^o$, a sharp decrease in both DOP$_{[110]}$ and DOP$_{[-110]}$ is noted due to the presence of hybridized H5 hole state around the center of the stack that contributes a drastic increase in the TM$_{[001]}$ mode. This is a promising result because it shows that DOP$_{[110]}$ and DOP$_{[-110]}$ can be simultaneously reduced below zero, leading to an isotropic polarization response from both [110] and [-110] cleaved-edge surfaces. This kind of polarization response is not accessible from the conventional [001]-aligned QDSs that only provide isotropic emission either from the [110] cleaved-edge~\cite{Inoue_1, Usman_1, Ikeuchi_1} or from the [-110] cleaved-edge\cite{Usman_7}.        

\begin{figure}
\includegraphics[scale=0.34]{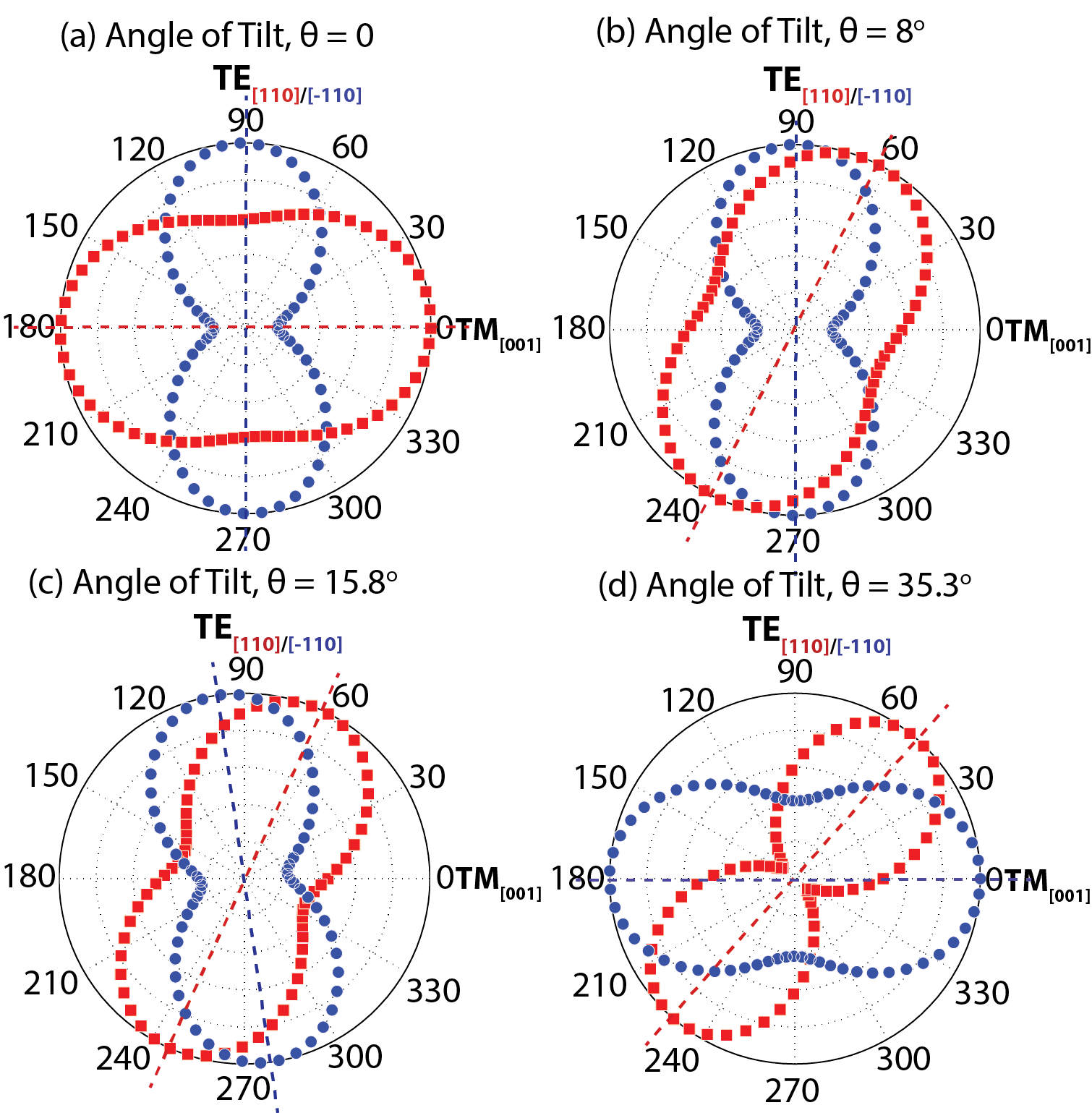}
\caption{Polar plots are shown for (a) $\theta$ = 0, (b) $\theta$ = 8$^o$, (c) $\theta$ = 15.8$^o$, and (d) $\theta$ = 35.3$^o$. The circles are for the plots obtained by varying the angle between the [001] and the [-110] directions, and the squares are for the plots obtained by varying the angle between the [001] and the [110] directions. All plots are normalized to 1.0.}
\label{fig:Fig8}
\end{figure}

The measured PL intensity polar plots showed a considerable difference between the angle of their peaks and the angle of tilt of the QDS obtained from TEM images~\citep{Bessho_1}. This inconsistency was associated with the difference between the quantum confinement direction and the strain orientation. Our calculations have shown highly asymmetrical nature of the biaxial strain profiles for the tilted QDSs, so to investigate its role on the measured discrepancy, we provide calculated polar plots (see Fig.~\ref{fig:Fig8}) for the QDSs with (a) $\theta$ = 0, (b) $\theta$ = 8$^o$ (c) $\theta$ = 15.8$^o$, and (d) $\theta$ = 35.3$^o$. The plots with the square symbols are drawn by varying the angle ($\theta_{p1}$) between the [001]-axis and the [110]-axis representing the PL intensity accessible from the [-110] cleaved-edge. The plots with the circle symbols are drawn by varying the angle ($\theta_{p2}$) between the [001]-axis and the [-110]-axis, illustrating the PL intensity attainable from the [110] cleaved-edge. All plots are normalized to 1.0.

For the [001]-aligned QDS, the PL intensity becomes maximum at $\theta_{p1}$ = 0 and $\theta_{p2}$ = 90$^o$, 
indicating that TM$_{[001]}$ is dominant from the [-110] cleaved-edge and TE$_{[-110]}$ is dominant from the [110] cleaved-edge. As the QDS is tilted, the highly asymmetrical biaxial strain tends to change the orientations of the hole wave functions towards the [110] edge of the QDLs (see Fig.~\ref{fig:Fig6}). At $\theta$ = 8$^o$ and 15.8$^o$, only one (H3) and two (H1 and H2) hole wave functions respectively are oriented along the [110]-direction and therefore the polar graphs for TE$_{[-110]}$ are peaked at $\theta_{p2} \approx$ 90$^o$ and $\approx$ 97$^o$ respectively, indicating that the TE$_{[-110]}$ mode is still dominant from the [110] cleaved-edge. However, the increase in the TE$_{[110]}$ leads to a large rotation of the [-110] cleaved-edge polar graphs, now exhibiting their peaks at $\theta_{p1} \approx$ 60$^o$ and $\approx$ 68$^o$ respectively in (b) and (c). These results are in qualitative agreement with the experimental report~\cite{Bessho_1}, where the peak of the PL intensity observed from the [-110] cleaved-edge surface showed a large rotation, whereas the peak of the PL intensity measured from the [110] cleaved-edge surface only slightly rotated. At $\theta$ = 35.3$^o$, all the hole wave functions are oriented along the [110]-direction leading to TE$_{[-110]}$ becoming much less than TM$_{[001]}$, and therefore the corresponding polar plot is peaked at $\theta_{p2}$ = 0 exhibiting a 90$^o$ rotation of the PL peak measured from the [110] cleaved-edge surface as a consequence of tilting the QDS. The TE$_{[110]}$ and TM$_{[001]}$ are however of nearly equal magnitudes at $\theta$ = 35.3$^o$ leading to the peak of the [-110] cleaved-edge polar plot to be at $\theta_{p2}$ $\approx$ 45$^o$. Overall the comparison of the polar plots clearly demonstrates that tilting of the InAs QDSs provides a high degree of control over their polarization response measured from both the [110] and [-110] cleaved-edge surfaces.  
   
In conclusion, we have studied electronic and polarization-dependent optical properties of the [110]-tilted QDSs by performing multi-million atom stacking-angle dependent simulations of the strain, electronic structure, and inter-band optical transition strengths. Our calculations reveal highly asymmetrical distributions of the biaxial strain components for the tilted QDSs, that govern the confinements and orientations of the hole wave functions thereby controlling the polarization properties. The computed values of the DOP$_{[110]}$ and DOP$_{[-110]}$ are in good agreement with the available experimental values at $\theta$ = 0 and 15$^o$. Our calculations predict that at $\theta$ = 35.3$^o$, isotropic polarization response could be simultaneously achieved from both the [110] and [-110] cleaved-edge surfaces, which has not been previously realized from the conventional [001]-aligned InAs/GaAs QDSs. A large rotation of the PL intensity peaks from the [-110] cleaved-edge surface indicates a strong impact of the asymmetrical strain profiles on the linear-polarization feature. Overall, the unique characteristics of the tilted QDSs highlight a novel technique to manipulate their polarization properties for the realization of several optoelectronic devices such as SOAs, lasers, intermediate-band solar cells, photo-detectors, etc.    

The author gratefully thanks Yukihiro Harada and Takashi Kita (Kobe University Japan) for useful discussions about their experimental data, and Stefan Schulz (Tyndall National Institute Ireland) for providing valuable feedback on the manuscript. The use of computational resources from nanoHub.org and Network for Computational Nanotechnology Purdue University USA are acknowledged.

\bibliographystyle{apsrev}       

\end{document}